\documentclass[fleqn,usenatbib]{mnras}

\usepackage{newtxtext,newtxmath}

\usepackage[T1]{fontenc}

\DeclareRobustCommand{\VAN}[3]{#2}
\let\VANthebibliography\thebibliography
\def\thebibliography{\DeclareRobustCommand{\VAN}[3]{##3}\VANthebibliography}

\usepackage{graphicx}	
\usepackage{amsmath}	
\usepackage{graphicx}
\usepackage{lscape}
\usepackage{color}
\usepackage{amsmath}
\usepackage{multirow}
\usepackage{booktabs}
\usepackage{caption}
\usepackage{subcaption}
\usepackage{ulem} 
\usepackage{parskip}
\usepackage{makecell}

\graphicspath{{./}{Figures/}}

\newcommand{\HI}{\hbox{\rmfamily H\,{\textsc i}}}
\newcommand{\HIsub}{\hbox{{\scriptsize H}\,{\tiny I}}}
\newcommand{\MHI}{\hbox{$M_{\HIsub}$}}
\newcommand{\msun}{\hbox{${\rm M}_{\odot}$}}
\newcommand{\kms}{\hbox{km\,s$^{-1}$}}
\newcommand{\askapsoft}{ASKAP{\sc soft}}

\title[DINGO Deep Imaging]{Deep Investigation of Neutral Gas Origins (DINGO): Options for robust Deep Spectral Line Imaging in the SKA-Era}

\author[J. Rhee et al.]{
Jonghwan Rhee$^{1,2}$\thanks{E-mail: Jonghwan.Rhee@csiro.au}, Richard Dodson$^{1}$, Alexander Williamson$^{1}$, Martin Meyer$^{1,3}$, Krist\'of Rozgonyi$^{1,3}$,
\newauthor Pascal J. Elahi$^{4}$, Matthew Whiting${^5}$, Daniel Mitchell$^{5}$, Tobias Westmeier$^{1,3}$ and Shinna Kim$^{1,3}$\\
\\
$^{1}$International Centre for Radio Astronomy Research (ICRAR), University of Western Australia, 35 Stirling Hwy, Crawley, WA 6009, Australia \\
$^{2}$Australia Telescope National Facility, CSIRO Space \& Astronomy, P.O. Box 1130, Bentley, WA 6102, Australia \\
$^{3}$ARC Centre of Excellence for All Sky Astrophysics in 3 Dimensions (ASTRO 3D), Australia \\
$^{4}$Pawsey Supercomputing Research Centre, Kensington, WA 6151, Australia\\
$^{5}$Australia Telescope National Facility, CSIRO Space \& Astronomy, P.O. Box 76, Epping, NSW 1710, Australia
}

\date{Accepted XXX. Received YYY; in original form ZZZ}

\pubyear{\the\year{}}

\begin{document}
\label{firstpage}
\pagerange{\pageref{firstpage}--\pageref{lastpage}}
\maketitle

\begin{abstract}
The data storage requirements for deep spectral line observations with next-generation radio interferometers like the Australian Square Kilometre Array Pathfinder (ASKAP) and the Square Kilometre Array (SKA) are challenging. The default strategy is to reduce data after each daily observation and stack the resulting images. Although computationally efficient, this approach risks propagating systematic errors (e.g. RFI, continuum and deconvolution residuals) and degrades data quality. Imaging the entire deep dataset jointly, the traditional approach, is prohibitively expensive in storage and compute. We present an alternative \textit{uv}-grid stacking method and compare its outcomes with both the traditional approach, our benchmark, and the default image-stacking method, using 200~h of the Deep Investigation of Neutral Gas Origins (DINGO) pilot and main survey data. Our method pauses the standard imaging pipeline after forming the daily residual visibility grids, which are then stacked and jointly deconvolved to combine many epochs of data. Relative to the traditional method, image-stacking recovers a median of 0.92$_{-0.02}^{+0.05}$ of the reference {\HI} flux across our source sample, and \textit{uv}-grid stacking recovers 0.99$_{-0.04}^{+0.01}$. For the brightest source, both methods show a similar, negligible flux offset of $\sim$3 per~cent from the traditional flux. {\HI} velocity widths ($W_{50}$, $W_{20}$) are recovered to within a few per~cent by both methods, with image-stacking showing somewhat larger deviations and scatter. Image-stacking further introduces non-physical artefacts, such as negative bowls around strong sources, indicating poor deconvolution and loss of physical information. Based on these findings, we intend to apply \textit{uv}-grid stacking to the DINGO survey on ASKAP.\\
\end{abstract}

\begin{keywords}
radio lines: galaxies -- methods: data analysis -- techniques: interferometric.
\end{keywords}



\section{Introduction}
New cutting-edge radio instruments such as the Square Kilometre Array \citep[SKA,][]{Dewdney:2009} will usher in a golden era of radio astronomy. The SKA will feature high-density interferometric arrays with hundreds or hundreds of thousands of individual antennas to enable wide-field, high-sensitivity, and high-angular resolution imaging of the radio sky.
However, to fully exploit this instrument, the radio astronomy community must manage and process its unprecedented volumes of data \citep{Quinn:2015, Scaife:2020}.

The SKA will be constructed in Australia, operating between 50 and 350~MHz \citep[SKA-Low,][]{Labate:2022}, and in South Africa, covering the frequency range of 350~MHz to 15.4~GHz \citep[SKA-Mid,][]{Swart:2022}.
In its first phase, SKA-Low will have 512 aperture array stations of 256 log-periodic dipole antennas each, deployed over a 42-m diameter, while SKA-Mid will consist of 133 offset Gregorian dishes with 15-m diameter and 64 MeerKAT dishes each of 13.5-m diameter. About 50~per~cent of stations/dishes will be located within a dense core area of $\sim$1-2~km, and the remaining are distributed along three spiral arms. The final goal is to have a full square kilometre of collecting area for both arrays.
Construction has commenced, and science verification will begin in 2027\footnote{\url{https://www.skao.int/en/647/timeline-science}}.

Following this, the data rates for each correlator will become $\sim$6\,Tb~s$^{-1}$ \citep{McMullin:2020}, making storage one of the largest cost drivers for the SKA project. 
Due to limited storage, raw data will be temporarily captured into a local buffer and must be processed within a specific period, ranging from days to weeks, before being permanently erased. 
This limitation will inevitably require real-time science data processing (SDP) with only a subset of raw or processed data being archived and provided for scientific use.
Thus, for multi-epoch projects, there will be no opportunity to iterate over all the data in the imaging process in the traditional approach, leading to one of the most outstanding challenges for the SKA--how to optimally image multi-day deep datasets. This limitation has the potential to cause all multi-epoch observations to be sensitivity-limited due to systematic errors, thus not achieving the scientific goals.

To address the unprecedented data volumes expected in the SKA era, various data compression strategies have emerged \citep[e.g.,][]{Pence:2009, Pence:2010, Offringa:2026, Gong:2023}. These range from general scientific data compression frameworks, such as those discussed by \citet{Cappello:2025}, to specialised applications for astronomy. Notable examples include Dysco for the lossy compression of noisy observed data \citep{Offringa:2016, Chege:2024} and Sisco for the lossless compression of forward-predicted model data \citep{Offringa:2026}. Furthermore, \citet{Gong:2023} introduced MGARD, a multigrid framework providing high-performance, error-controlled compression and refactoring. Early evaluations of MGARD in radio astronomy indicate it can compress visibility data by approximately a factor of 20 without introducing structural artefacts \citep{Gong:2023, Dodson:2025}.

Alongside these bit-level technologies, Baseline-Dependent Averaging (BDA) has been proposed to reduce interferometric visibility volumes \citep{Cotton:1986, Cotton:2009, Wijnholds:2018, Atemkeng:2016, Atemkeng:2018, Atemkeng:2023, Deng:2022}. BDA exploits the principle that the rate at which a baseline tracks through the \textit{uv}-plane is proportional to its length; thus, shorter baselines can be averaged more aggressively than longer ones without significant loss of information. Similarly, Sidereal Visibility Averaging (SVA), proposed by \citet{de_Jong:2025} and motivated by the work of \citet{Owen:2008} and \citet{Owen:2018}, leverages the repeating baseline tracks inherent in deep, multi-epoch observations. By averaging calibrated data from different sidereal days at identical baseline coordinates, SVA reduces the total number of visibilities, thereby lowering the computational burden and long-term storage costs for ultra-deep imaging.

Finally, \citet{Golap:2016} proposed using gridded visibilities as a compressed format for combining multi-year observations (e.g., the CHILES \citep{Fernandez:2013} project), implemented as the CASA \citep{CASA-Team:2022} task \texttt{msuvbin}. Unlike BDA, this method accumulates data onto a fixed, uniform \textit{uv}-grid to mitigate time and bandwidth smearing. This approach serves as the foundation for the \citet{Rozgonyi:2021} thesis, which addressed original limitations in the \texttt{msuvbin} model, specifically regarding anti-aliasing, wide-field (non-coplanar) imaging effects, and preconditioning schemes. The study presented in this paper builds directly upon this gridded visibility model.

The SKA precursors are good testbeds for preparing for the forthcoming SKA era by demonstrating scientific and technical readiness. The Australian Square Kilometre Array Pathfinder \citep[ASKAP,][]{Johnston:2007, Johnston:2008, Hotan:2021} plays a key role in addressing the technical challenges that large-scale radio survey projects of the SKA will face. 
The Deep Investigation of Neutral Gas Origins \citep[DINGO,][]{Meyer:2009a, Rhee:2023} is one of the ASKAP survey science projects and aims to carry out deep spectral-line ({\HI}) observations (3,200~h integration in total) over the G23 region of the Galaxy And Mass Assembly \citep[GAMA,][]{Driver:2011, Hopkins:2013, Liske:2015, Baldry:2018, Driver:2022} survey.
This region has the richest multi-wavelength datasets, enabling both {\HI} and continuum science, and establishing a legacy dataset of maximal value for SKA-era surveys. 
The {\HI} data from DINGO will yield some of the deepest images ever taken of atomic hydrogen gas content in the Universe, enabling groundbreaking new studies of the role that this fundamental component has played in the ongoing evolution of galaxies and its connection to their dark matter halos, in conjunction with multi-wavelength data from GAMA. 

On top of its scientific merits, the DINGO data will be beneficial to test and demonstrate a solution for the SKA technical challenges in deep spectral line imaging and long-term data storage, as the DINGO survey will collect spectral line data spread over many epochs to obtain an 800~h integration per tile. The final raw data volume would be $\sim$ 13.2~PB, which is beyond what is feasibly stored. In this paper, we explore three deep spectral line imaging methodologies with 200~h of data from the DINGO pilot and main surveys, while a companion paper \citep{Williamson:2024} focuses on the performance tests we undertook to prove that our \textit{uv}-gridding method is a solution for long-term data storage in radio deep survey projects.

The paper is structured as follows. Section~2 gives details of the DINGO data used in this work, e.g., the target field, observations, and initial data processing. In Section~3, we describe three deep imaging methodologies compared in this paper. Section~4 presents the results of a comparison between three deep imaging methods, followed by our discussion and conclusions in Section~5 and 6.
We adopt the concordance cosmological parameters \citep{Hinshaw:2013} of $\Omega_{\Lambda}=$~0.7, $\Omega_{M}=$~0.3 and $H_{0}=$~70~km~s$^{-1}$~Mpc$^{-1}$ throughout this paper.

\section{Data}
\subsection{Observations}
\begin{figure}
    \centering
    \includegraphics[width=\linewidth]{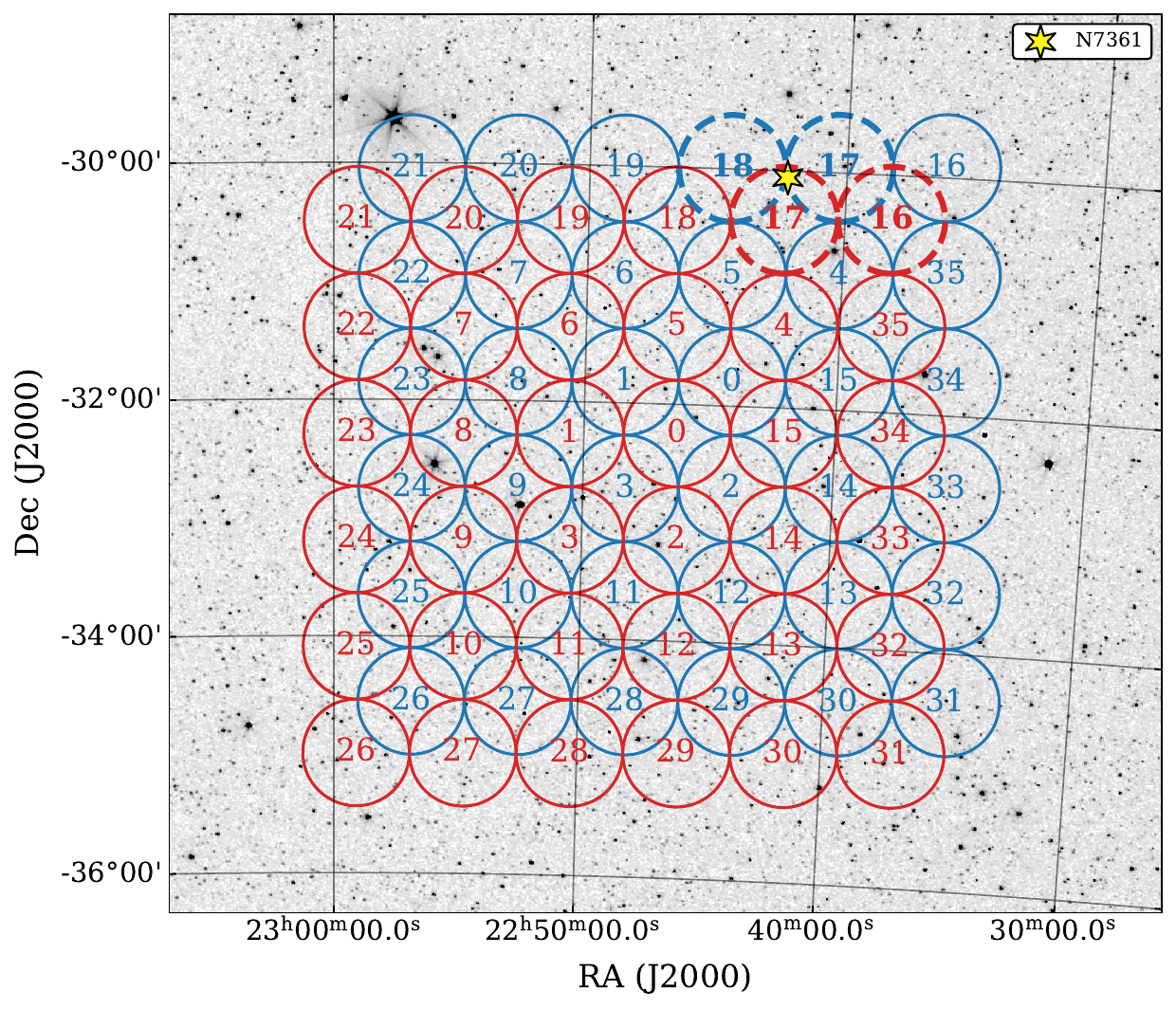}
    \caption{ASKAP beam footprints, with diameters of 0.9${\degr}$ each, used for DINGO observations over one tile of the G23 region. Two 36-beam patterns (blue for footprint A and red for footprint B, respectively) of each 6-by-6 beam configuration are interleaved to achieve the uniform sensitivity over the field. Beam IDs are overlaid on individual beams. For this study, we select two beams (dashed-line circles) from each footprint, surrounding a bright {\HI} galaxy of NGC~7361 (denoted with a yellow star) already detected in this field by the HIPASS survey \citep{Koribalski:2004}. The background is the optical all-sky image from \citet{Mellinger:2009}.}
    \label{fig:footprint}
\end{figure}

This work used data taken from the DINGO pilot and main survey observations over the G23 field, corresponding to 25 scheduling blocks (SBs) and approximately 200~h integration time in total (see Table~\ref{tab:obs_summary}). 
The DINGO pilot observations were made in two phases between 2019 and 2022, while the main survey observations began in 2023. We carried out DINGO pilot and main survey observations with the full ASKAP array of 36 antennas and 288~MHz bandwidth split into of 15,552 channels of 18.52~kHz channel width, corresponding to a velocity resolution of $\sim4~\kms$ at $z\sim0$. The integration time of a normal DINGO scheduling block (SB) is about 8~h except for some of the pilot phase 1 observations where shorter observations were conducted. The ASKAP phased array feeds (PAFs) receiver generates a squared 6~$\times$~6 (\texttt{square\_6$\times$6}) beam configuration with an angular separation of 0.9${\degr}$ between adjacent beams, forming an instantaneous field of view (FoV) of $\sim$~30~deg$^2$. The entire G23 field of 12$\times$5 deg$^2$ is divided into two tiles, and each tile is covered by two interleaved \texttt{square\_6$\times$6} footprints to achieve uniform sensitivity over the tile, as seen in Figure~\ref{fig:footprint}. The pilot and main survey observations taken so far focused on one tile in the G23 region, centred at $\alpha, \delta$ (J2000) = $22^{\rm h}47^{\rm m}17^{\rm s}78, -32\degr29\arcmin03\farcs26$.

In pilot phase 1, the two interleaved pointings were alternated every 15 minutes to have one SB contain two footprints to cover an entire tile at a time. However, from pilot phase 2, we changed this observing strategy to have one SB with one footprint. This change led to reducing the flagged data fraction and increasing the integration time of each footprint by a factor of two for deeper imaging.
Along with science data, calibration data were obtained by observing PKS~B1934-638 before or after science observation. 

\begin{table*}
\centering
  \caption{The summary of DINGO pilot and main survey observations.}
  \label{tab:obs_summary}
  \begin{tabular}{lcccccccc}
    \toprule
    Survey & SBID & Date & Obs. Time & Footprint$^a$ & No. of Beams & \makecell{Data Volume \\ per beam$^b$} & Bandwidth & Centre Frequency \\
           &      &      & [hr]      &               &              & [GB]                              & [MHz]     & [MHz]            \\
    \midrule
    Pilot Phase 1 & 10991 & 2019-12-25 & 8.15 & A/B & \multirow{7}{*}{36} & 232 & \multirow{7}{*}{288} & \multirow{7}{*}{1295.5} \\
                  & 10994 & 2019-12-26 & 6.15 & A/B &                     & 176 &                      &                         \\
                  & 11000 & 2019-12-28 & 8.16 & A/B &                     & 232 &                      &                         \\
                  & 11003 & 2019-12-29 & 8.82 & A/B &                     & 251 &                      &                         \\
                  & 11006 & 2019-12-30 & 8.17 & A/B &                     & 232 &                      &                         \\
                  & 11010 & 2019-12-27 & 5.70 & A/B &                     & 162 &                      &                         \\
                  & 11026 & 2020-01-02 & 8.17 & A/B &                     & 232 &                      &                         \\
    \midrule
    Pilot Phase 2 & 39697 & 2022-04-20 & 8.02 & A & \multirow{6}{*}{36} & 456 & \multirow{6}{*}{288} & \multirow{6}{*}{1295.5} \\
                  & 32214 & 2021-09-18 & 8.04 & B &                     & 456 &                      &                         \\
                  & 39722 & 2022-04-21 & 8.03 & A &                     & 456 &                      &                         \\
                  & 39995 & 2022-04-28 & 8.02 & B &                     & 456 &                      &                         \\
                  & 40196 & 2022-05-03 & 8.03 & A &                     & 456 &                      &                         \\
                  & 40018 & 2022-04-29 & 8.03 & B &                     & 456 &                      &                         \\
    \midrule
    Main          & 65177 & 2024-08-26 & 8.05 & A & \multirow{12}{*}{36} & 456 & \multirow{12}{*}{288} & \multirow{12}{*}{1295.5} \\
                  & 65210 & 2024-08-27 & 8.03 & A &                      & 456 &                       &                          \\
                  & 65268 & 2024-08-28 & 8.03 & A &                      & 456 &                       &                          \\
                  & 65297 & 2024-08-29 & 8.03 & A &                      & 456 &                       &                          \\
                  & 65347 & 2024-08-30 & 8.03 & A &                      & 456 &                       &                          \\
                  & 65405 & 2024-08-31 & 8.03 & A &                      & 456 &                       &                          \\
                  & 65665 & 2024-09-06 & 8.03 & B &                      & 456 &                       &                          \\
                  & 65690 & 2024-09-07 & 8.03 & B &                      & 456 &                       &                          \\
                  & 65748 & 2024-09-11 & 8.03 & B &                      & 456 &                       &                          \\
                  & 65818 & 2024-09-13 & 8.03 & B &                      & 456 &                       &                          \\
                  & 65836 & 2024-09-14 & 8.04 & B &                      & 456 &                       &                          \\
                  & 65853 & 2024-09-15 & 8.10 & B &                      & 456 &                       &                          \\
    \bottomrule
    \multicolumn{9}{l}{$^a$Pilot phase 1 observations alternated between footprint A and B in one SB.}\\
    \multicolumn{9}{l}{$^b$ Since SBs of Pilot phase 1 observations have both footprint A and B in one SB, the data volume here is per-beam and per-footprint.}\\
  \end{tabular}
\end{table*}

\subsection{Data Reduction}
\label{sec:data_reduction}
The DINGO data reduction pipeline processes each science SB with {\askapsoft} \citep{Cornwell:2011, Whiting:2017, Wieringa:2020}, ASKAP's science data processing  software\footnote{\url{https://www.atnf.csiro.au/computing/software/askapsoft/sdp/docs/current/index.html}}, to produce DINGO spectral line datasets used for this work, following a standard {\HI} data reduction procedure. This includes flagging, bandpass/flux calibration, self-calibration, continuum subtraction, and spectral line imaging. More details on DINGO data reduction are described in \citet{Rhee:2023}. 
These data reduction steps were applied separately to individual beams in each daily SB.
The wide-spread RFI impact of global navigation satellite systems (GNSS) below 1295.5 MHz restricts us to using only the upper half of the 288-MHz bandwidth, spanning 1295.5 to 1439.5 MHz, corresponding to 7776 channels.

For subsequent deep spectral line imaging analysis, we used calibrated and continuum-subtracted spectral line visibility data of four beams around a well-known bright galaxy \citep[NGC~7361,][]{Koribalski:2004} in {\HI} (see Figure~\ref{fig:footprint}). After imaging processes of the three methods compared in this work, an additional continuum subtraction was performed in the image domain to remove continuum residuals. Then, we combine four beams using the mosaicking command of {\askapsoft}, \texttt{linmos}. This mosaicking process includes correction for the ASKAP primary beam response using the ASKAP holography model \citep{Hotan:2021} with the threshold set at a 20 per~cent cut-off of the peak response.

\section{Method}
\label{sec:method}
In this study, we explore three deep imaging methods to identify an optimal spectral line imaging strategy for DINGO, and also establishing a foundation for future deep SKA surveys. Using identical processing parameters across all methods, we analysed the same dataset via: (1) the traditional approach (`visibility stacking'), where visibility data from all observations are co-added prior to deep imaging; (2) the current default methodology, commonly known as `image stacking' \citep[e.g.,][]{Knudsen:2015}, for deep projects with ASKAP and the SKA, where data are imaged by scheduling block (approximately daily observations) and the resulting images are co-added with a weight per pixel representing the exposure in the image domain, where no deep CLEANing is possible; and (3) a novel approach (`grid stacking') that we developed, where processing halts after forming \textit{uv}-grids of the daily data at the final major deconvolution cycle, with these residual grids subsequently stacked with weights per \textit{uv}-cell before performing a final minor deconvolution cycle, followed by the restoration of the model components into the
image cube, convolved with a Gaussian beam that matches the resolution. 

\begin{figure*}
    \centering
    \includegraphics[width=0.9\linewidth]{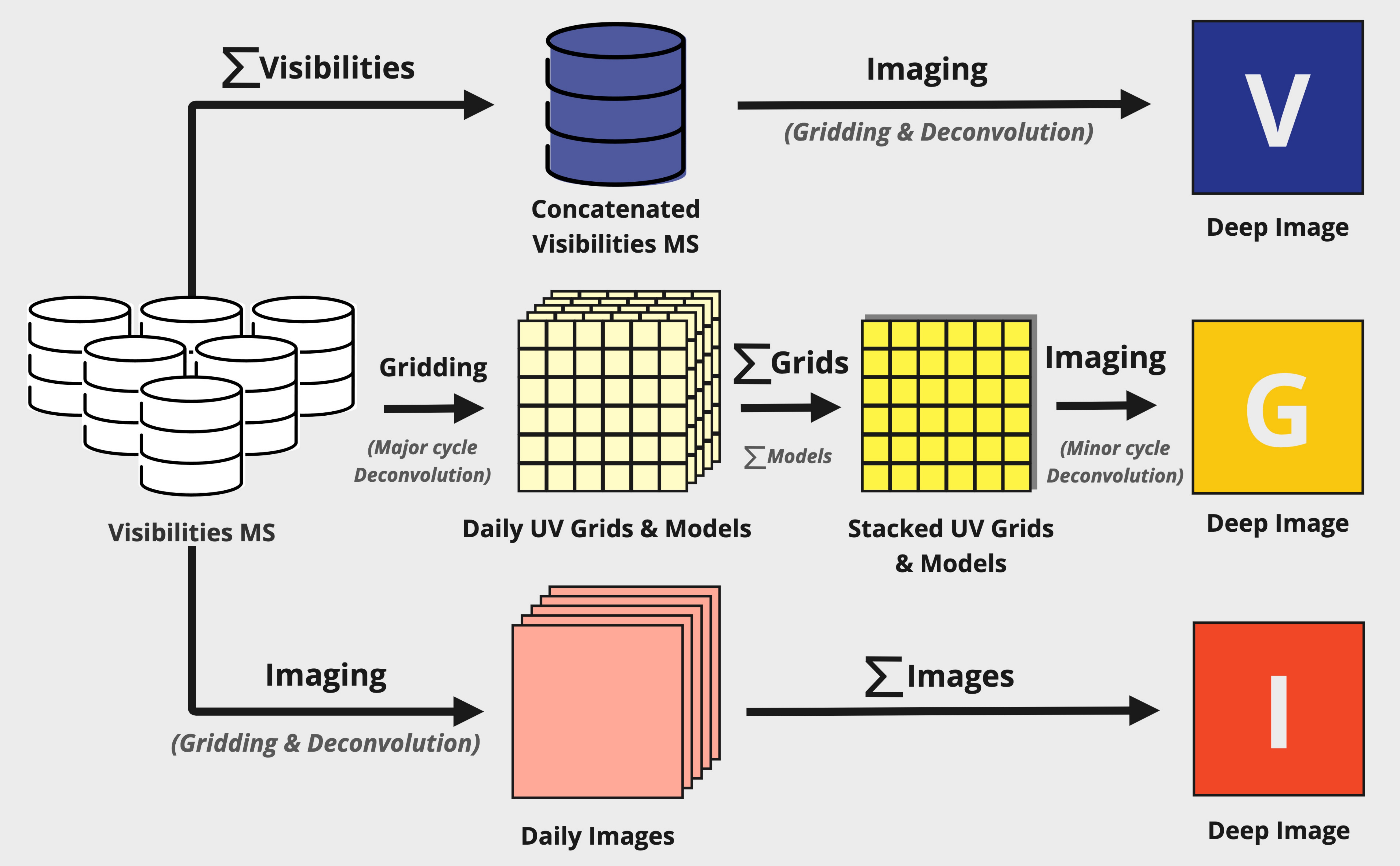}
    \caption{Schematic flowchart illustrating the three deep imaging strategies compared in this study using DINGO data. From top to bottom, the panels represent visibility stacking (V), grid stacking (G), and image stacking (I). Visibility stacking follows the standard approach of concatenating all spectral-line Measurement Sets (MS) prior to imaging. Grid stacking conceptually decouples the gridding and deconvolution processes by generating residual \textit{uv}-grids and models for each individual observation; these are then combined to undergo an additional minor-cycle deconvolution. The final G image is produced by combining the stacked models (convolved with the restoring beam) with the deconvolved residual grids. Image stacking entails the direct co-addition of individual image cubes processed via the standard ASKAPsoft pipeline. The $\Sigma$ symbol denotes the stacking stage for each method.} 

    \label{fig:methods}
\end{figure*}

The schematic workflows of these deep imaging approaches are illustrated in Figure~\ref{fig:methods}. In all cases, data reduction was performed with very similar parameters; four overlapping beams (except the image stacking method that used all beams) with an image size of 1024$\times$1024 per beam, with a robustness parameter of 0.5 and additional Gaussian tapering to achieve a restored beam size of $\sim$~30\arcsec{} across the frequency channels and a cell size of 6\arcsec, and maximum baselines limited to 2\,km.

The next step is to combine all the processed beams using the {\askapsoft} mosaicking process as described in section~\ref{sec:data_reduction}. We then conduct {\HI} source finding on the mosaicked data products from the three methods and compare physical properties of {\HI} sources commonly detected across the datasets, especially focusing on NGC~7361, the well-known {\HI} source in the field. For this analysis, we use an automated 3D source finding software dedicated to {\HI} surveys, called Source Finding Application \citep[SoFiA,][]{Serra:2015a, Westmeier:2021}. It produces moment maps and {\HI} spectra of detected sources with their physical parameters derived, such as positions (or coordinates), {\HI} velocity (or redshift), fluxes (peak and integrated), and velocity widths (e.g., $W_{20}$ and $W_{50}$). 
In the following subsections, we describe the details of the respective workflows.

\subsection{Traditional Visibility Stacking}
The traditional approach for imaging multi-epoch data, the most developed and best understood methodology, is to combine the multiple measured visibility data in the $\textit uv$ domain, which is then imaged with the CLEAN algorithm \citep{Hogbom:1974, Clark:1980}, one of the most successful deconvolution and restoration procedures in radio astronomy \citep{Cornwell:2009}. The CLEAN processing of multi-epoch data combined in the $\textit uv$ domain is robust against artefacts caused by low-level continuum residuals, RFI and/or systematic errors lurking in shallower-depth data. In this regard, the traditional method is taken as our baseline, against which we can compare the results from the other methods. However, it also has the most challenging resource requirements, with long-term storage of the observed visibility data products currently prohibitive for both SKA and ASKAP. In light of this limitation, the observatory policy was to adjust the survey strategy to reduce each day independently and sum those images, allowing for the deletion of the observed visibilities.

Given the current data volume of the DINGO survey, it is still technically feasible to test this traditional deep imaging approach for a limited number of beams. To this end, we combined 16 SBs of DINGO pilot and main survey data for each footprint, corresponding to $\sim$~5.6~TB of spectral-line measurement sets (MSs) for the same beam and footprint, as seen in Figure~\ref{tab:obs_summary}. Instead of physically concatenating 16 MSs followed by imaging, {\askapsoft} allows users to simultaneously execute these two steps in its imaging step. The {\askapsoft} imaging command, called \texttt{imager}, accepts multiple MSs of the same beam and footprint as input, enabling both combination and deconvolution during the imaging procedure. Consequently, we define `visibility stacking' as the process of concatenating multiple measurement sets in time-order and gridding each dataset onto a common plane prior to the CLEAN deconvolution process. This term specifically refers to the superposition of gridded visibilities on a shared plane, rather than the binning or averaging of individual visibility data points. Using this functionality, we imaged four beams' data using \texttt{imager} with its CLEAN algorithm optimised for ASKAP processing, \textit{BasisfunctionMFS}, a re-implemented and improved version of the CASA's multi-scale deconvolution algorithm \citep{Cornwell:2008} with \textit{w}-projection \citep{Cornwell:2008a}, \textit{w}-stacking \citep{Offringa:2016} and \textit{w}-snapshot \citep{Cornwell:2012} algorithms available to account for non-coplanar baseline effects due to \textit{w} terms.

After all the combined data has been gridded on a regular grid in the 2D $\textit (u, v)$ space per frequency slice, it is Fourier-transformed to produce an image, called dirty image. Sky-models are created by iteratively identifying peaks in that image from which the instrumental sampling response function (i.e., the Point Spread Function, PSF, the FFT of the weights convolved with the gridding kernel) is subtracted from this dirty image. This is the so-called minor-cycle CLEAN. The major cycle CLEAN involves degridding those models, reprojecting them onto the original regular time- and frequency-sampled visibility data, and subtracting the models at these points, after which the residuals are regridded, imaged, and passed through the minor cycle again. Refer to \citet{Thompson:2017} for more details about the CLEAN algorithm.

For all methods discussed in this paper, the CLEAN process performed three major cycles, with minor cycles configured for a maximum of 1000 iterations per major cycle. This setup aligns with the standard ASKAP {\HI} data reduction pipeline, further optimised for the DINGO survey specifications. The process typically executes until the residual emission reaches the designated threshold.
Specifically, we applied a general threshold of 5$\sigma$ across the entire field without a mask, supplemented by a deep CLEAN down to 0.5$\sigma$ within auto-masked regions, which are defined by pixels where CLEAN components were previously identified. {\askapsoft} allows these threshold parameters to be defined either dynamically on a noise basis (as a multiple of the localised noise level) or statically in units of Janskys ($\text{Jy}$). To ensure a rigorous comparative analysis, we applied identical imaging parameters, including the noise-based CLEAN thresholds, iteration limits, and the number of major cycles, across all three methods.

As mentioned in Section 2.2, the input spectral line MSs used are already continuum-subtracted in the \textit{uv}-domain at the depth of the daily data, rather than at the full 200~h level. However, we performed additional continuum subtraction from the data cubes of each beam to remove residual continuum emission by subtracting a second-order polynomial fit to the spectrum at each pixel in the data cubes. After this step, we combined the four beams' cubes independently processed using a linear mosaicking command of {\askapsoft}.

\subsection{Image Stacking}
\label{sec:imstack_method}
The DINGO data are processed on a daily basis (SB) using the {\askapsoft} pipeline with DINGO-specific processing parameters. This image-stacking method combines individual image cubes deconvolved and restored by the pipeline in the image domain. Following calibration and \textit{uv}-domain continuum subtraction, three major-cycle CLEAN iterations are performed with the same noise-based thresholds used for the visibility stacking method during spectral line imaging. This is followed by a secondary image-domain continuum subtraction, consistent with the traditional approach. 

The spectral line data cubes from all 36 beams in each SB are then combined using the {\askapsoft} mosaicking command. This stage substantially reduces the volume of the data products relative to the input visibility data, as the 36 individual beam overlap at their half-power points. Such a reduction is critical, given that ASKAP (and other instruments such as SKA) does not intend to retain raw visibility data after the imaging products are generated. Finally, the data cubes from 16 SBs per footprint are co-added into a final data cube covering the entire field of interest, as seen in Figure~\ref{fig:footprint}. 

Since this processing is done on the daily subset of the entire data, it is computationally straight forward and inexpensive, and necessary for rapid data validation. However, this approach has strong disadvantages in that it irreversibly embeds all artefacts deeper than the individual observational level, leaving sidelobe, continuum residual artefacts, and low-level RFI that could appear more obviously in accumulated data. The vast majority of final spectral line sources from this method at the long integration level (800~h per tile and band for the DINGO main survey) is unlikely to be CLEANed appropriately.

\subsection{Grid Stacking}
Given the limitations of the two approaches mentioned above, a new intermediate approach was needed and we have adopted the stacking of the \textit{uv}-grid residuals. Our preferred solution is to store the daily residual data in the \textit{uv}-grid along with CLEANed model data from major cycles and daily Point Spread Function (PSF) data, which can reduce storage requirements over the traditional approach. 
This method was first introduced as the \texttt{msuvbin} tool in CASA \citep{Golap:2016}. It has been further developed as part of the PhD thesis of \citet{Rozgonyi:2021} and the shared ICRAR-Pawsey-AusSRC HiVIS\footnote{\HI\ Visibility Imaging for the SKA} PaCER\footnote{Pawsey Centre for Exascale Readiness} project \citep{Rozgonyi:2022}. 
It has the advantage of enabling minor-cycle CLEANing at full survey depth and improving continuum subtraction.  
However, since the daily data are stored after application of the \textit{uv}-kernels and combined on the 2D \textit{uv}-plane, major cycles are not possible, so we must ensure that we work with nearly final residuals, so that minor-cycle processing is able to adequately deconvolve the full-depth data.
Although this is a highly promising approach and the data products of this method can be stored long-term (enabling future reprocessing), detailed comparison of this approach to the other deep imaging solutions has not been published until now, other than in the PhD thesis of \citet{Rozgonyi:2021}. We note that while we have adopted a method of stacking residual \textit{uv}-grids in this paper, a similar deep data product could be achieved by stacking residual images. However, this approach does not offer the same data storage/processing savings and potential future re-processing advantages, as discussed in Section~\ref{discuss:residual-stacking}.

The generation of daily residual grids has been integrated into the standard ASKAP DINGO processing using the \texttt{imager} tool with minimal computational overhead, provided the cubes are CLEANed with three major cycles. To optimise storage, the {\askapsoft} Nyquist gridding option was employed, which trims the zero-padding from the \textit{uv}-grid before they are stored. This reduces the typical grid size to $\sim$~600~$\times$~600 pixels. Additionally, the {\askapsoft} Pre-Convolution Function (PCF)--a count of the visibilities prior to the application of kernels and tapering--must be retained for grid-stacking to facilitate proper renormalisation. 

However, several additional processing steps are required for this method.
Firstly, the daily \textit{uv}-grid residuals, PSF and PCF--which are required for an additional deconvolution iteration--must be stacked and then deconvolved using the {\askapsoft} deconvolution command, \texttt{cdeconvolver}.
Subsequently, the daily CLEAN models are summed and convolved with the restoring (`clean') beam. These components are then added back into the final data products once the final residual cube is generated, following a procedure similar to that used for the final model components, albeit as a separate processing step. Prior to mosaicking cubes of the individual beams generated via grid-stacking, an additional continuum subtraction step is performed to maintain consistency with the other processing methods.

\section{Results}
\begin{figure*}
    \centering
    \includegraphics[width=1.0\linewidth]{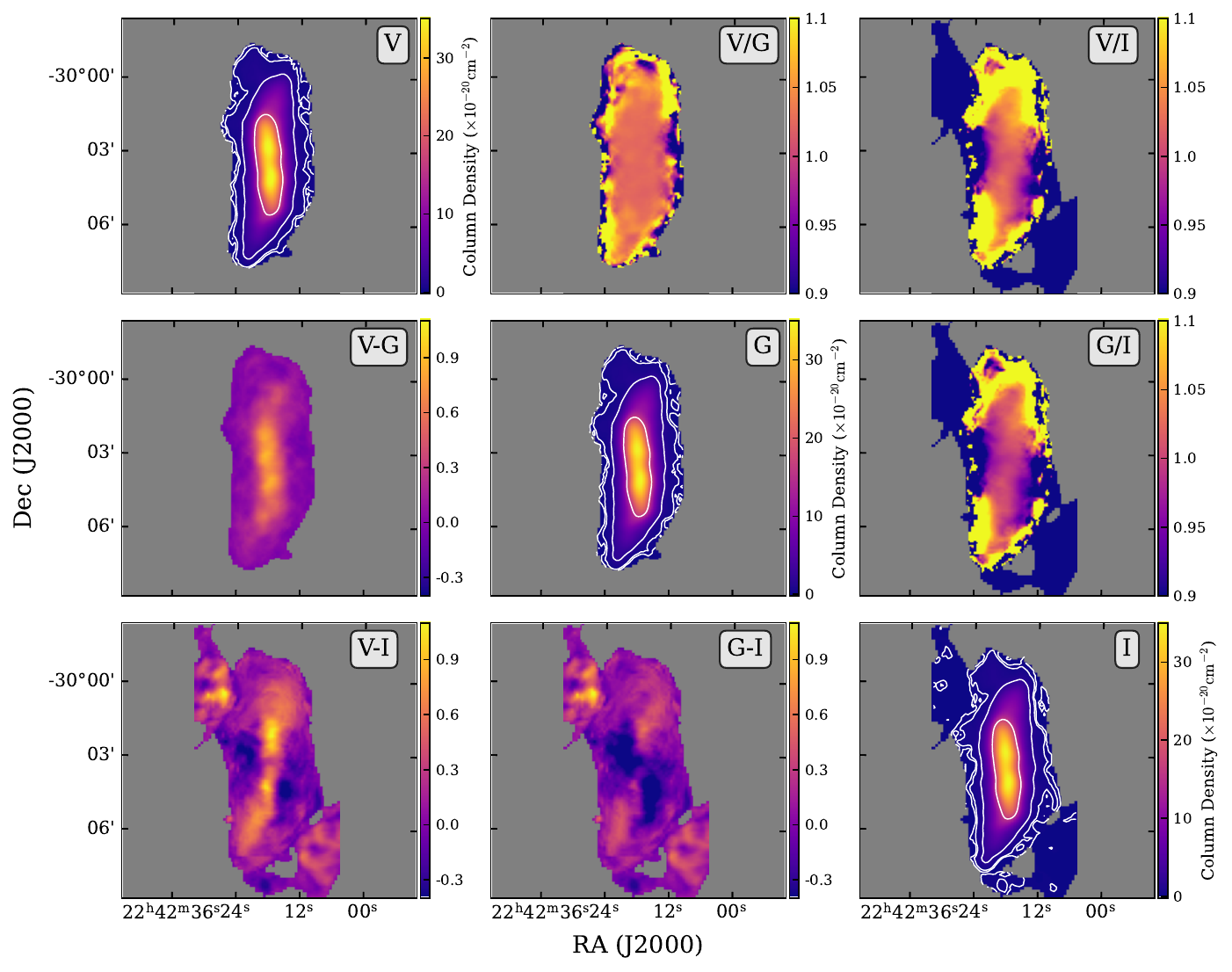}
    \caption{Comparison of the moment 0 images made with the three methods for NGC~7361. Shown on the diagonal are the moment 0 maps converted to column density for traditional visibility stacking (V), grid stacking (G) and image stacking (I), respectively.
    The lower off-diagonal plots are the differences between these three approaches and the 
    upper off-diagonal ones are the ratios. Whilst the three approaches give very similar looking results, it is obvious that low level residuals remain in the image-stacking case, whereas the other two methods give very similar results. This is born out in the difference shown in V-G (visibility stacking minus grid stacking image) where there is a small ($\sim$2\%) residual, wholly contained within the region of the galaxy with maximum flux. In comparison, the difference shown in V-I (visibility stacking minus image stacking) contains negative and positive features, and is not aligned well with the original image. The ratios reaffirm this result, in that the ratio of the V/G is smooth over the area with {\HI} emission whereas the ratio of V/I shows a great deal of structure and a much greater range of values. 
    \label{fig:mom0-maps}}
\end{figure*}

\begin{figure}
    \centering
    \includegraphics[width=1.0\linewidth]{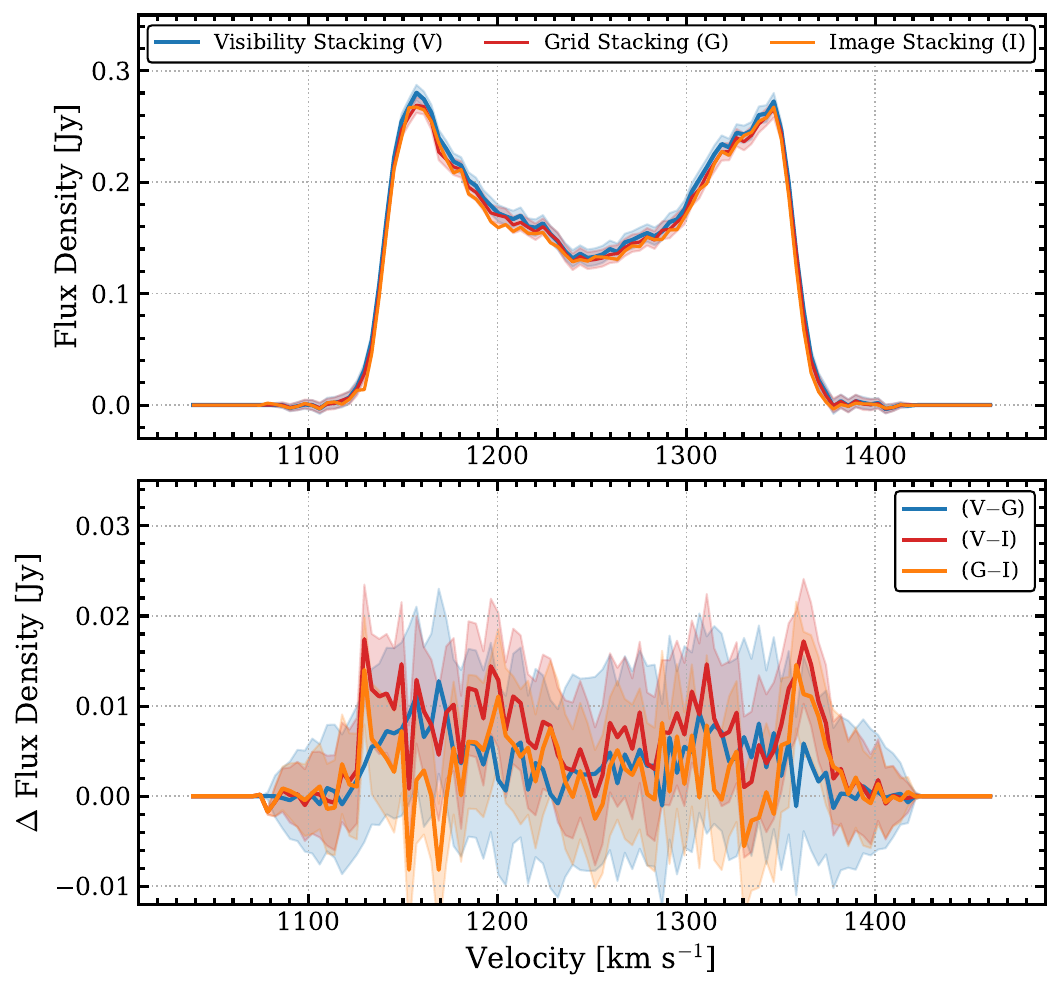}
    \caption{Comparison of the {\HI} spectra made with the three different methods (V, G and I) for NGC~7361 (top). The bottom panel shows the differences between three spectra. The shaded area denotes the uncertainty of each {\HI} and residual spectrum.
    Grid-stacking residuals are close to a constant fraction of the V profile, which suggests the cleaning cutoff needs to be deeper in this method where there are no major cycles. Image-stacking results have more complex structure, which suggests there is no simple remedy. 
    }
    \label{fig:spectra}
\end{figure}

\subsection{Comparison of the Images}
\label{subsec:compare_image}
The comparison of the three imaging pipelines demonstrates the benefits and disadvantages of the different strategies, where we treat the traditional visibility-stacking image as close to the `ground truth' as we can obtain. 
For the daily images, the residual level was the same regardless of whether images or grids were produced, as their generation process is identical. For stacked images, the residual level is lower and very similar between the different methods, as detailed in Table~\ref{tab:ngc7361_params}. 

In Figures~\ref{fig:mom0-maps} and \ref{fig:spectra} we compare the images from the three pipelines with their differences and ratio, and the derived {\HI} spectra for the strongest source in the DINGO observation field, NGC~7361.
In Figure~\ref{fig:mom0-maps} on the diagonal are the moment 0 images of the three methods (V, traditional visibility-stacking; G, grid-stacking; I, image-stacking), converted to {\HI} column density. Below the diagonal, we show the differences between the methods, while above the diagonal, we show the column density ratios. Both are masked to include only regions covered by the SoFiA source mask. The contour levels of 1, 7, 50 and 500~$\sigma$ column density sensitivities are overlaid for reference. 

All three methods yield broadly similar results. However, the image-stacking moment map shows that there are noticeable and extended residual sidelobes left around the galaxy, likely due to the limited deconvolution capability inherent to the method. The difference in the comparison of the moment 0 maps is greater for the V-I image than for the V-G image, with peak differences approximately 26\% higher. Additionally, the ratio of V/I shows a greater range and more complex structures than the ratio of V/G. In Figure~\ref{fig:spectra}, we also plot the spectra integrated over the source in the images and their respective differences. The spectral differences are at the few-percent level, with the V-G difference echoing the original spectra, while the V-I difference shows greater scatter and slightly higher residuals. This is consistent with the results presented in Figure~\ref{fig:mom0-maps}.

Our results clearly demonstrate the limitations of the default ASKAP image-stacking method, particularly its failure to accurately reproduce the traditional visibility-stacking results around strong sources, as shown in Figure~\ref{fig:mom0-maps}. This is not unexpected; previous work has shown the importance of retaining visibility data in some form for deep spectral line imaging \citep[e.g., as discussed in][]{Dodson:2016}. 

Stacking daily images also would co-add uncleaned residuals into the final product. Consequently, sidelobes from uncleaned flux, especially around strong sources, can dominate the final image, as seen in Figure~\ref{fig:mom0-maps}. Our study, however, shows that grid-stacking achieves results broadly comparable to those obtained with visibility-stacking regarding spatial fidelity in the image domain. Given that the storage required for full visibility-stacking is not feasible for large-scale surveys, grid-stacking presents a demonstrated and viable solution.

\begin{figure*}
    \centering
    \includegraphics[width=0.9\linewidth]{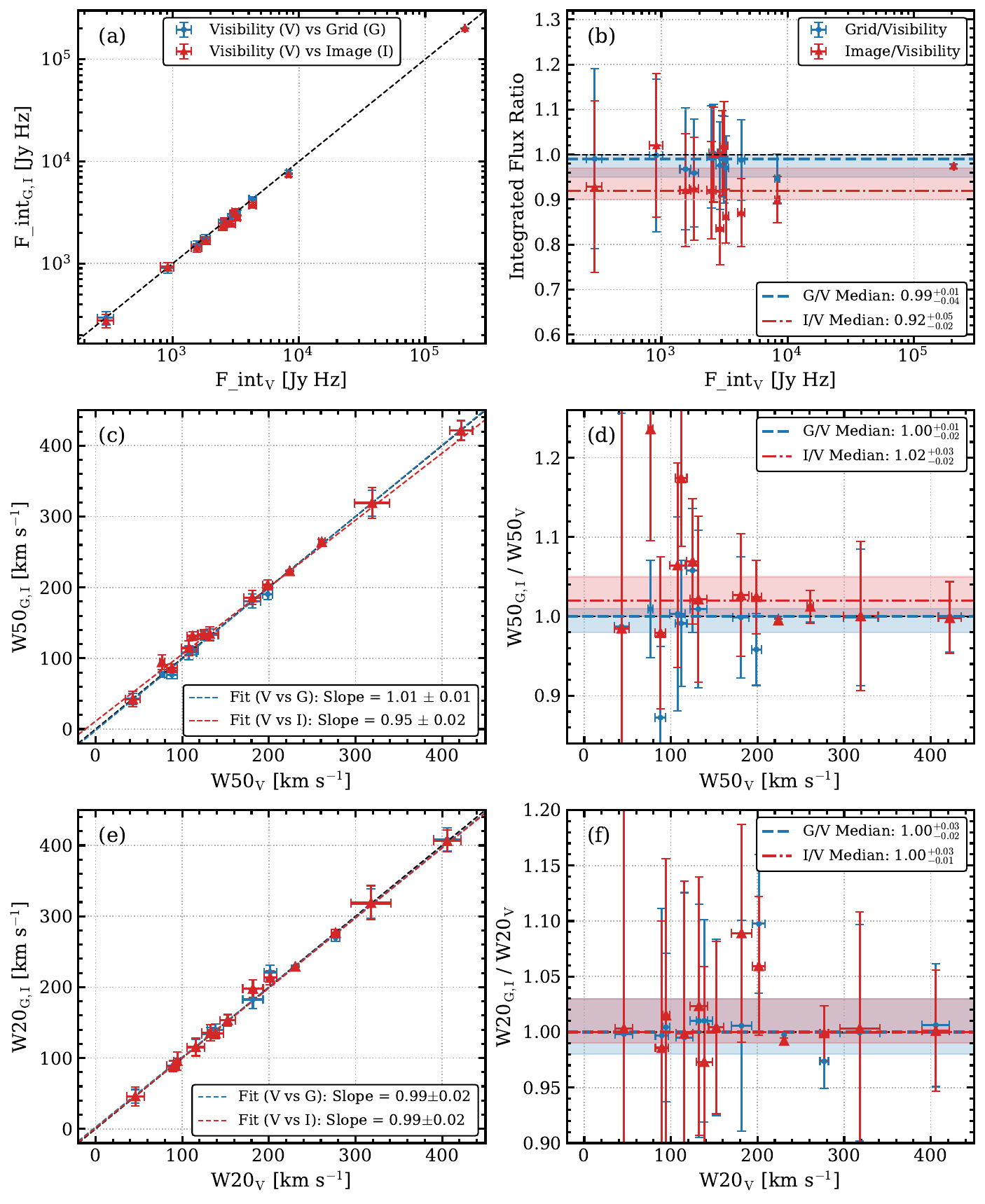}
    \caption{The comparison of derived physical parameters for detected sources using traditional visibility stacking (V), grid-stacking (G, blue circles) and image-stacking (I, red triangles). (a) Integrated flux: integrated flux from grid stacking (F\_int$_{\rm G}$) and image stacking (F\_int$_{\rm I}$) on y-axis versus the traditional visibility stacking (F\_int$_{\rm V}$) on x-axis. (b) Integrated flux ratio: The ratio of the stacked integrated flux (Grid/Visibility in blue; Image/Visibility in red) over the traditional integrated flux. The median ratios are indicated by the dashed lines: 0.99$_{-0.04}^{+0.01}$ for grid-stacking (blue) and 0.92$_{-0.02}^{+0.05}$ for image-stacking (red), respectively. The shaded areas denote the 1~$\sigma$ uncertainties of the median flux ratios. (c) The {\HI} line width at 50\% of the mean flux density across the spectrum ($W_{50}$) derived from grid-stacking and image-stacking (y-axis) against the traditional visibility stacking result (x-axis). Dashed lines show linear fits, with slopes of 1.01$\pm$0.02 (blue, V vs G) and 0.95$\pm$0.02 (red, V vs I). (d) The ratio of $W_{50}$ from grid-stacking and image-stacking $W_{50_{\rm G,I}}$/$W_{50_{\rm V}}$) over the traditional visibility stacking method with the 1~$\sigma$ uncertainty range. The median ratios are 1.00$_{-0.02}^{+0.01}$ (blue dashed line and shade, G/V) and 1.02$_{-0.02}^{+0.03}$ (red dashed line and shade, I/V). (e) Same as (c), but for the {\HI} line width at 20\% of the peak flux ($W_{20}$). Fitted slopes are 0.99$\pm$0.01 for both V vs G (blue) and V vs I (red). (f) Same as (d), but for the $W_{20}$ line width ratio and its uncertainty ($W_{20_{\rm G,I}}$/$W_{20_{\rm V}}$). The median ratios are 1.00$_{-0.02}^{+0.03}$ (blue dashed line, G/V) and 1.00$_{-0.01}^{+0.03}$ (red dashed line, I/V), respectively.
    }
\label{fig:results-HI}
\end{figure*}

\subsection{Comparison of the Physical Parameters}
\label{subsec:compare_physical_paras}
\begin{table}
\caption{Comparison of the physical parameters of NGC 7361, brightest {\HI} source in the field, derived from SoFiA runs on the three images generated by the different methods. The close agreement between them confirms that the three methods are equivalent in the strong signal domain.}
\label{tab:ngc7361_params}
\addtolength{\tabcolsep}{-4pt}
\begin{tabular}{lccc}
\toprule
Parameter & V & G & I \\
\midrule
RMS  [mJy~beam$^{-1}$] & 0.64 & 0.64 & 0.64 \\
S$_{\rm int}$ [Jy~Hz] & 206427.4$\pm$701.0 & 200995.4$\pm$700.5 & 201087.4$\pm$724.0 \\
$W_{20}$ [{\kms}] & 230.68$\pm$0.42 & 230.08$\pm$0.44 & 228.87$\pm$0.45 \\
$W_{50}$ [{\kms}] & 224.07$\pm$0.31 & 223.65$\pm$0.31 & 223.00$\pm$0.34 \\
{\rm log}$_{10}${\MHI} [{\msun}] & 9.52$\pm$0.32 & 9.51$\pm$0.31  & 9.51$\pm$0.31 \\
\bottomrule
\end{tabular}
\end{table}
In Table~\ref{tab:ngc7361_params}, we compare the physical properties of NGC~7361 derived from the three images shown in Figure~\ref{fig:mom0-maps}. All three datasets exhibit identical noise characteristics, with SoFiA providing the RMS of the local noise within each source's bounding box. The integrated flux for both grid-stacking and image-stacking is $\sim$2.6~per~cent lower than the traditional (visibility-stacking) value, while the $W_{50}$ and $W_{20}$ velocity profile widths differ from visibility-stacking by $\lesssim$1 per~cent for both methods. Although the formal per-source uncertainties in Table~\ref{tab:ngc7361_params} are small enough to resolve these offsets from zero, their absolute size remains negligible, and we regard the grid-stacking and image-stacking results for this single, high signal-to-noise ratio (SNR) source as consistent with the traditional benchmark in a practical sense, while noting that grid-stacking and image-stacking are not clearly distinguished from one another at this SNR level.

Figure~\ref{fig:results-HI} presents the physical properties of the 13 detected {\HI} sources from the combined four-beam subset of data. The figure compares results from the visibility-stacked (V), image-stacked (I), and grid-stacked (G) images. The upper left panel shows the integrated fluxes measured by SoFiA from each image cube, with the image-stacked and grid-stacked fluxes plotted against the traditional visibility-stacked fluxes. The upper right panel shows the ratio of these fluxes against the traditional measurements. The lower panels display the {\HI} line widths, specifically $W_{50}$ and $W_{20}$, defined as the line widths at 50\% and 20\% of the mean flux density and the peak flux density, respectively. A robust method based on \citet{Courtois:2009} was used to measure $W_{50}$ to mitigate the effects of noise (see the details in the SoFiA manual\footnote{\url{https://gitlab.com/SoFiA-Admin/SoFiA-2/-/wikis/documents/SoFiA-2_User_Manual.pdf}}). 
We estimated the uncertainties for $W_{50}$ and $W_{20}$ by means of Monte Carlo simulations. First, each detected integrated {\HI} spectrum was modelled using the busy function, an analytic profile specifically designed for {\HI} line profile \citep{Westmeier:2014}. We then generated 1,000 realisations of the spectrum by injecting Gaussian random noise, scaled to the per-channel flux uncertainties derived from the local RMS of the original data, into the best-fit model spectrum. Each realisation was refitted with the busy function and the resulting interquartile range (IQR) of the $W_{50}$ and $W_{20}$ distributions was adopted as the formal uncertainties for each measurement. The ratio of $W_{50}$ and $W_{20}$ are plotted in the lower right panels. 

The integrated fluxes both from the image-stacked and grid-stacked data are generally consistent with the visibility-stacked measurements, as shown in panel (a) of Figure~\ref{fig:results-HI}. However, the integrated flux ratios presented in panel (b) reveal a systematic offset between the two methods; the median flux ratios for image-stacking and grid-stacking are 0.92$_{-0.02}^{+0.05}$ and 0.99$_{-0.04}^{+0.01}$, respectively. Uncertainties for the median values were estimated using bootstrap resampling with 1,000 iterations, where $1\sigma$ error ranges were derived from the 16th and 84th percentiles of the resulting distribution. This disparity indicates that the grid-stacking method recovers the flux more reliably on average across the sample.

This sample-wide discrepancy contrasts with the comparable flux offsets found for the brightest source alone (NGC~7361; Table~\ref{tab:ngc7361_params}), where both grid-stacking and image-stacking underestimate the traditional value by a similar $\sim$2.6 per cent, despite image-stacking showing visually the more prominent residual sidelobe structure than the two methods (Figure~\ref{fig:mom0-maps}). This indicates that the amplitude of the residual artefact is not simply proportional to the resulting fractional flux bias, and instead points to an SNR-dependent effect. The flux left unrecovered by the shallower, daily-depth CLEAN used in image-stacking is expected to be of broadly similar absolute magnitude across sources in the same field, since the CLEAN threshold is set relative to a similar local noise level. Such a residual represents only a small fractional bias for a bright source such as NGC~7361, but a substantially larger one for the fainter sources that dominate the sample median in Figure~\ref{fig:results-HI}, consistent with the median image-stacking flux ratio of 0.92$_{-0.02}^{+0.05}$. Grid-stacking, by contrast, recovers a median of 0.99$_{-0.04}^{+0.01}$ of the traditional flux. We believe a deeper CLEAN on the \texttt{cdeconvolver} step might close the remaining gap, although this would break our initial experimental design, being that we used identical setups for as much of the parameters as possible.

The corresponding effect on the velocity widths is comparatively small. $W_{50}$ and $W_{20}$ are only marginally affected by either stacking method in our sample ($\lesssim$2 per~cent; panels (d) and (f)). The comparative fits for $W_{50}$ and $W_{20}$, as shown in panels (c) and (e) respectively, track each other consistently between the traditional, the image- and grid-stacked data. However, for the $W_{50}$ measurement, the relationship derived by comparing the traditional and image-stacking methods is of slightly lower quality. Specifically, the image-stacking results show a greater deviation from a 1:1 relationship than those obtained via the grid-stacking method. This deviation is evidenced by a $\sim$11\,\kms{} offset in the $W_{50}$ measurement fit for the image-stacked data, as opposed to a smaller $\sim$2\,\kms{} offset found with the grid-stacking method. The ratio plots of the velocity widths further underpin this finding. The median $W_{50}$ ratio for the image-stacked data is 2~per~cent higher than that of the traditional method, while the median ratio for the grid-stacked data is close to unity. 

\subsection{Comparison of the Computing Resources}
The imaging processing for all three methods (traditional visibility stacking, image stacking, and grid stacking) exploits parallelisation using the Message Passing Interface (MPI). This is achieved by dividing the 7,776 spectral line channels into 216 chunks of 36 channels each, and assigning them to 217 MPI processes (comprising 216 worker processes and one master process). These processes were distributed across 2 compute nodes on the Setonix supercomputer at the Pawsey Supercomputing Research Centre, with each node configured for a maximum of 109 tasks.

Table \ref{tab:results-full_resources} lists the computional requirements for the imaging processing with the three different pipelines. In addition to the spectral line imaging step, all methods require subsequent continuum subtraction in the image domain to remove residual continuum emission, and two stages of linear mosaicking. The first mosaicking step combines two beams from each footprint (36 beams for the full analysis) with primary beam response correction using the ASKAP holography model \citep{Hotan:2021}, followed by a second step that mosaics the resulting two footprint datasets to produce the final image cube. 

For a complete survey comprising both footprints, these post-processing steps must be applied to 72 beams. However, we focus our resource analysis on the spectral line imaging step, as this dominates the computational burden and also varies most between methods.

\subsubsection{Traditional Visibility Stacking}
Traditional visibility stacking has the highest instantaneous resource requirements of the three approaches, as it requires the simultaneous processing of the entire dataset. For one beam, we required approximately 10 CPU-hours. Crucially, the I/O requirements are dramatically elevated: reading the combined 5.6 TB dataset requires approximately 3,900 GB of I/O operations—roughly 100 times greater than expected from simple linear scaling of the 16 daily imaging runs. We attribute this excessive I/O overhead likely to cache limitations when handling such large data volumes, though this aspect requires further investigation.

The major cycle CLEAN limits were based on the same relative RMS threshold as for the daily imaging runs, but with 16 times more data, these limits translated to lower absolute flux thresholds, enabling deeper cleaning. Despite the high instantaneous requirements, this approach required approximately 9\% less total compute time than the cumulative sum of the 16 individual image-stacking processes. The memory footprint, while several times greater than image stacking at approximately 3-7 GB, remained manageable with this volume of data (200 h observation) for modern HPC systems.

\subsubsection{Image Stacking}
The image stacking method has the lowest per-epoch resource usage of the three approaches. Each of the 25 scheduling blocks (SBs), split into 16 for each footprint, for each 36-beam footprint was imaged independently with the same processing parameters (i.e. 3 major cycles, cleaning to 5$\sigma$, deep cleaning to 0.5$\sigma$ and 1,000 minor cycle iterations) as the other methods, then combined in the image domain. As expected, the per-epoch CPU time ($\sim$41 minutes per beam) was substantially shorter than traditional processing, and the memory footprint ($\sim$1.6 GB~average) was significantly lower than both alternatives.

However, the true computational cost emerges when considering all epochs: for 16 epochs, the total compute time per beam is 16 $\times$ 41.3 $\times$ 217 $\approx$ 2,400 CPU-hours, approximately 9\% higher than traditional processing. The primary advantage of this method lies in its distributed temporal nature—the compute burden is spread over the multi-year survey duration rather than concentrated at the end. Furthermore, the modest I/O requirements ($\sim$40 GB read, $\sim$0.6 GB write per epoch) make this approach operationally attractive for facilities with limited storage infrastructure. 

For the full DINGO survey, image stacking requires minimal long-term storage since only final image products must be retained, allowing processing to be distributed across the survey lifetime.

\subsubsection{Grid Stacking}
Grid stacking represents an optimised compromise between resource requirements and data quality. Like image stacking, 16 epochs of two beams per footprint were imaged independently on the Setonix supercomputer using 217 cores. The per-epoch compute time (80 minutes per beam) was about two times longer than image stacking likely because it produces both a CLEAN component model image and a residual \textit{uv}-grid. But this is an acceptable increase considering the significant improvement in data quality demonstrated in Section~\ref{subsec:compare_image} and \ref{subsec:compare_physical_paras}.

The key difference lies in the memory footprint: grid stacking requires approximately 29 GB peak memory (versus $\sim$4.5 GB for image stacking), a factor of $\sim$6.5 increase. This elevated memory usage stems from retaining and writing out the full residual \textit{uv}-grids along with the CLEAN model components. However, this remains well within the capabilities of modern HPC facilities. The I/O characteristics are similar to image stacking for write operations ($\sim$0.5 GB per epoch on average), but the average read operation slightly increased due to the need to access the full \textit{uv}-grid datasets.

Grid stacking requires an additional final processing step: after stacking the 16 residual \textit{uv}-grids per beam, a final minor-cycle deconvolution is performed using the ASKAPsoft \texttt{cdeconvolver} command. This step requires 41 minutes per beam with 217 CPUs--a trivial addition ($\sim$3\% of total compute) executed only once per beam after all observations are complete. The daily CLEAN models must then be summed, convolved with the 'clean beam', and added back to the final residual cube in a separate processing step, adding minimal computational overhead.

For the full DINGO survey with all 72 beams per tile, grid stacking requires storage of compressed \textit{uv}-grids ($\sim$7:1 compression ratio achievable, as discussed in \citet{Williamson:2024}) rather than raw visibilities, reducing the storage burden to manageable levels while enabling full-depth deconvolution.

\begin{table*}
    \caption{Reported compute resources used in the spectral line imaging processing (only) of one beam by the three methods, all of which processed 7776 channels of continuum subtracted spectral line data, with 3 major cycles per channel for one beam. Identical processing parameters were applied for every method, including the limit to 217 MPI processes in the slurm run.}
    \label{tab:results-full_resources}
    \centering
    \begin{tabular}{c|rrr|lr|ll|rr}
    \toprule
      Imaging  & epochs$\times$ &  time  & CPU  &  \multicolumn{2}{c|}{Memory [GB]} & \multicolumn{4}{c}{I/O [GB]} \\
      Method   & beams$\times$  & [mins] &   cores     &     Max.    &    Ave.       & \multicolumn{2}{c}{Read} & \multicolumn{2}{c}{Write} \\
               &                &        &           &           &             &  Max.  &  Ave. &  Max.  & Ave. \\
      (1) & (2) & (3) & (4) & (5) & (6) & (7) & (8) & (9) & (10) \\
      \midrule
      Visibility Stacking & 1$^\dagger\times$72$\times$ &  608   & $\times$217 & 6.82  & 3.16  & 3908  & 3850 & 0.60 & 0.58 \\
      Image Stacking      &     16$\times$72$\times$    &  41.3  & $\times$217 & 4.48  & 1.62  & 40.44 & 6.39 & 0.57 & 0.09 \\ 
      Grid  Stacking      &     16$\times$72$\times$    &  80.0  & $\times$217 & 29.02 & 3.26  & 39.72 & 39.53 & 27.22 & 0.50 \\ 
    \bottomrule   
    \end{tabular}
    \vspace{0.7em}
    \begin{minipage}{0.68\linewidth} 
    \footnotesize 
    \textit{Note}. Col~(1): Imaging method. Col~(2): The number of epochs and beams for full processing estimation. However, for this work only 4 beams were processed for the comparison. Col~(3): The average processing time in minutes for one beam. Col~(4): The number of MPI tasks, with one CPU core per task. Col~(5) and (6): Memory usage (Max. and Ave.). Cols~(7)--(10): I/O usage (Read Max./Ave. and Write Max./Ave.). \\
    \end{minipage}
    
\end{table*}

\section{Discussion}
\begin{figure}
    \centering
    \includegraphics[width=0.9\linewidth]{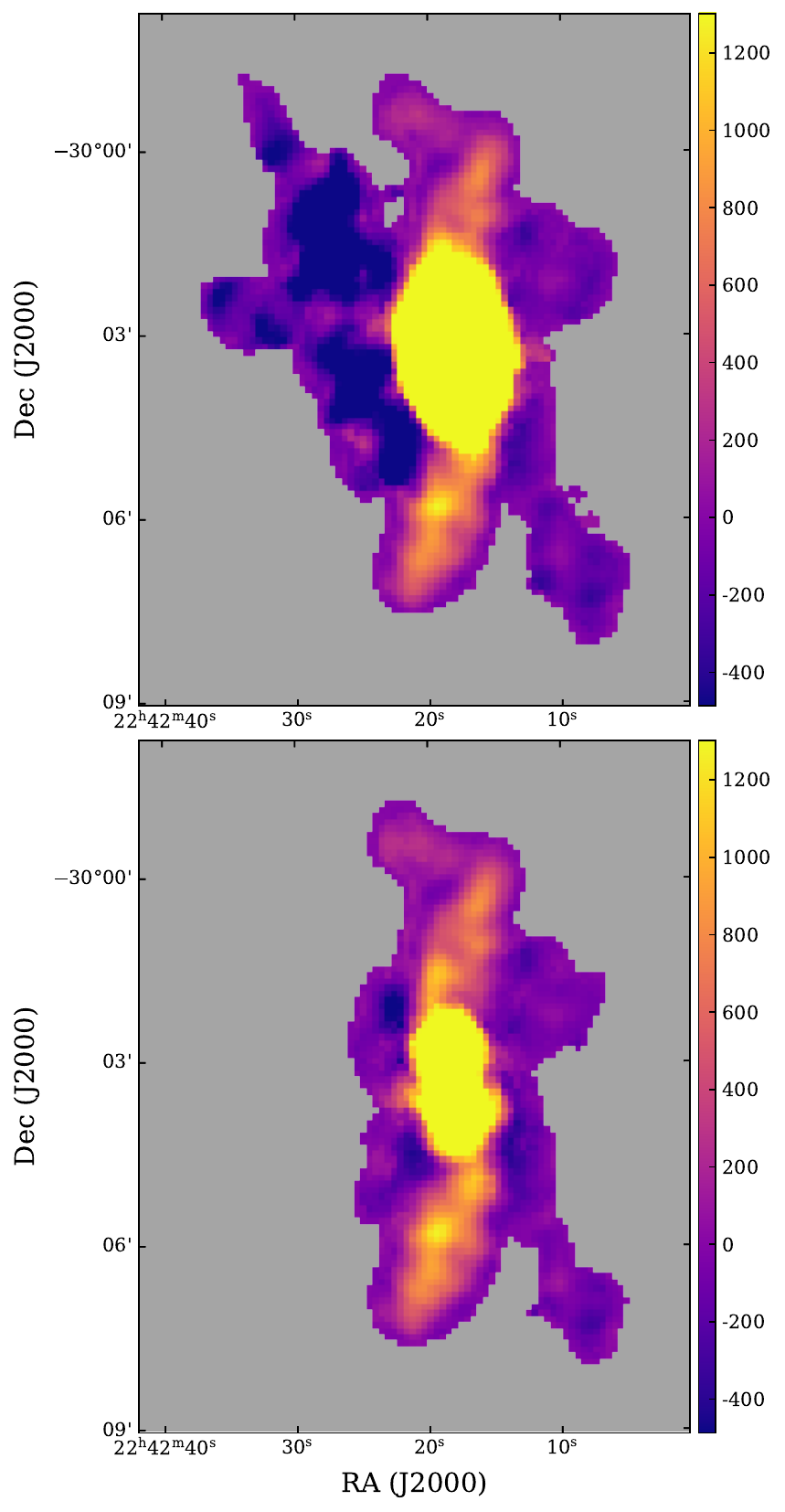}
    \caption{Comparison of NGC~7361 moment 0 maps before (\textit{top}) and after (\textit{bottom}) an additional minor-cycle deconvolution was applied to the stacked residual image of Beam 17. Prominent sidelobe artefacts are visible around the source prior to deconvolution---particularly on the eastern (\textit{left}) side of the galaxy---whereas these artefacts are successfully suppressed following the additional minor-cycle deconvolution stage.}   

\label{fig:residual_deconvolution}
\end{figure}

\subsection{Residual Image Stacking}
\label{discuss:residual-stacking}
We briefly tested an alternative approach to grid-stacking, termed ‘residual image-stacking’, where a minor-cycle CLEAN is performed directly on the deep residual image. This is conceptually similar to grid-stacking, where faint sidelobe artefacts are removed via an additional minor-cycle deconvolution step. The final stage common to grid- and residual image-stacking methods is the restoration of the model components into the final image cube, convolved with a Gaussian restoring beam that matches the resolution. The fundamental distinction between the two approaches lies in whether the stacking is executed in the \textit{uv} or the image domain. Notably, implementing residual image-stacking within a data reduction pipeline requires storing the daily image model, PSF, and residuals, in addition to the standard data products.

To evaluate whether residual image-stacking reproduces the results comparable to grid-stacking, we performed an additional deconvolution using the co-added residual and PSF images. Because {\askapsoft} does not currently support this functionality for image data (as opposed to gridded data), we utilised the CASA \texttt{deconvolve} task. We applied minor-cycle deconvolution parameters closely matching those used in our grid-stacking analysis presented in Section~\ref{sec:method}: a maximum of 1000 iterations, multi-scale CLEAN using the identical scales, and an equivalent CLEAN threshold.  

Figure~\ref{fig:residual_deconvolution} displays the moment 0 maps of NGC~7361 before and after this additional minor-cycle deconvolution (\textit{top} and \textit{bottom} panels, respectively). 
Sidelobe artefacts are prominent in the deep residual image prior to cleaning (\textit{top} panel), whereas the additional minor-cycle deconvolution successfully mitigates these artefacts as clearly seen on the eastern side of the galaxy in the bottom panel. This result indicates that grid- and residual image-stacking yield equivalent performance regarding sidelobe artefact suppression. However, a primary operational constraint is that neither the daily images nor the PSFs compress efficiently, because the image domain contains fewer truly empty pixels compared to gridded data products. Consequently, compression of these daily 32-bit floating-point image products yields a data volume reduction of less than a factor of 2 \citep{Pence:2009,Pence:2010}, compared to nearly an order of magnitude achieved for the \textit{uv} domain products \citep{Williamson:2024}.

\subsection{Compression}
In this work, we focus on a key comparison: whether or not the grid-stacking can deliver a final data product as good as that from the traditional processing and significantly better than image stacking approach. There are a number of other advantages for grid stacking that have not been covered in this paper.

We have not mentioned the impact of the compression on the \textit{uv}-grids as this is covered in the partner paper of \citet{Williamson:2024}. The total disk footprint of a \textit{uv}-grid is larger than the image domain transform, because it is complex and one also requires the matching full-sized PSF (sum of weights, convolved with the \textit{w}-kernels and tapering) and PCF (the count of samples used, which is required for the summation of the grids, as real values and kernel size as imaginary values). However, these grids compress very efficiently, as the values are largely zero (and integer in the case of the PCF). Compression ratios of $\sim$7:1 have been found for the \textit{uv}-grids, and higher if we allow for lossy compression \citep{Williamson:2024}. The reduced size of the data products directly speeds up the processing time in I/O limited tasks. This is critical as next-generation instruments are considered IO-bound.

\subsection{Future work}
Because radio frequency interference (RFI) is typically limited in both time and frequency and is often confined to shorter baselines, its characteristics can be used to identify weak, persistent RFI sources. By comparing \textit{uv}-cells across multiple epochs, we can identify these contaminated cells and mitigate weak RFIs before the final imaging stage. \texttt{msuvbin} of CASA has demonstrated this application with \texttt{msuvbinflag} \citep{Golap:2016}. We have also tested the feasibility of identifying RFI-affected regions within the data cubes; however, a comprehensive analysis of this approach is beyond the scope of this paper and left for future work.

If the \textit{uv}-grids were also sub-divided on the \textit{w}-planes, one could implement major cycle CLEAN. Whilst this would increase dramatically the in-memory footprint of the grids, it would not much alter the compressed size, as the empty cells compress efficiently. 
Alternatively, the \textit{uv}-grids could be sub-divided by baseline, which would also allow for solution of the \textit{w}-terms, and also allow for station-based recalibration. 
Again, this would dramatically increase the in-memory footprint of the grids, but it would not considerably alter the storage footprint of data products.

\section{Summary and Conclusions}

We have compared three methods for deep spectral line imaging suitable for next-generation radio instruments, using 200~h of data from the DINGO pilot and main survey. This multi-epoch dataset, spanning 16 days for each footprint, was combined and imaged using three distinct pipelines: traditional visibility stacking, image stacking, and grid stacking.
Traditional visibility stacking, which involves jointly imaging all stored visibility data, is not technically feasible for the full SKA or even ASKAP deep surveys due to the enormous disk space required. While image stacking, the current default for ASKAP and the SKA due to its computational efficiency, co-adds daily spectral line images, this method carries the critical risk of propagating systematic effects, particularly deconvolution errors. As a demonstrated solution, we developed the grid-stacking method for the DINGO survey, which co-adds residual grids from daily imaging before a final deconvolution.

Our comparison, using traditional processing (visibility stacking) as a benchmark, shows a picture for grid-stacking overall, with some caveats. Averaged across our sample of 13 sources, image stacking recovers only 0.92$_{-0.02}^{+0.05}$\% of the expected {\HI} flux, compared to 0.99$_{-0.04}^{+0.01}$\% for grid stacking, which is a clear, sample-wide improvement, though this advantage is less pronounced for the single brightest source in our field (NGC~7361), where the two methods perform comparably (Table~\ref{tab:ngc7361_params}). Image stacking also shows systematically larger deviations and scatter than grid stacking in the {\HI} velocity profile widths, particularly $W_{50}$, and further introduces marked non-physical artefacts, including negative bowls and extended residual sidelobes around the strong source (see Figure~\ref{fig:mom0-maps}). These artefacts arise from the irreversible embedding of uncleaned residuals from daily imaging into the final product. Grid stacking achieves results that more closely match traditional processing overall, because it enables minor-cycle deconvolution at full survey depth while avoiding the propagation of daily-level systematic errors. Similar results have been found for MeerKAT spectral line data (Dodson et al. 2026, submitted)

The computational resource analysis reveals that traditional visibility stacking requires the highest instantaneous resource and $\sim$1.6~PB of storage for the full DINGO survey with 800~h and 72 beams per tile--operationally infeasible for ASKAP and impossible for the SKA. Image stacking, however, requires minimal storage but produces degraded data quality. Grid stacking provides the reasonable balance: it requires only twice the compute time per epoch than image stacking while achieving substantially better flux recovery. The elevated memory footprint remains well within modern HPC capabilities. More importantly, efficient compression of the \textit{uv}-grids ($\sim$7:1 ratio) reduces storage requirements to manageable levels, while preserving the ability to perform full-depth deconvolution.

Based on these findings, we will implement the \textit{uv}-grid stacking method for the full DINGO survey on ASKAP. Grid stacking represents a practical and demonstrated improvement over the current default approach for one of the most pressing technical challenges facing the SKA era. Our results with 200~h of DINGO data provide strong evidence that this method can deliver the data quality closer to the traditional benchmark than the current default, supporting its use for transformative {\HI} science in the next decade.

\section*{Acknowledgements}
We are grateful to the anonymous referee for helpful comments that have improved this paper. This scientific work uses data obtained from Inyarrimanha Ilgari Bundara / the Murchison Radio-astronomy Observatory. We acknowledge the Wajarri Yamaji People as the Traditional Owners and native title holders of the Observatory site. CSIRO’s ASKAP radio telescope is part of the Australia Telescope National Facility (https://ror.org/05qajvd42). Operation of ASKAP is funded by the Australian Government with support from the National Collaborative Research Infrastructure Strategy. ASKAP uses the resources of the Pawsey Supercomputing Research Centre. Establishment of ASKAP, Inyarrimanha Ilgari Bundara, the CSIRO Murchison Radio-astronomy Observatory and the Pawsey Supercomputing Research Centre are initiatives of the Australian Government, with support from the Government of Western Australia and the Science and Industry Endowment Fund. This work was supported by resources provided by the Pawsey Supercomputing Centre under project code ja3 with funding from the Australian Government and the Government of Western Australia. Part of this research was supported by the Australian Research Council Centre of Excellence for All Sky Astrophysics in 3 Dimensions (ASTRO 3D) through project number CE170100013. Part of this research was supported by the Pawsey PACER project. 
This research made use of Astropy, a community-developed core Python package for Astronomy \citep{Astropy-Collaboration:2013}, Matplotlib \citep{Hunter:2007}, and Numpy \citep{Harris:2020}.

\section*{Data Availability}
The data underlying this article will be shared upon reasonable request to the corresponding author, JR.




\bibliographystyle{mnras}
\bibliography{references_dingo_deep_imaging} 








\bsp	
\label{lastpage}
\end{document}